\documentclass[pre,twocolumn,amssymb,showpacs,superscriptaddress,notitlepage]{revtex4-1}
\pdfoutput=1
\usepackage{graphicx}
\usepackage{color}
\usepackage{epsfig}
\usepackage{latexsym}
\usepackage{bm}
\usepackage{ulem}
\usepackage{color}
\usepackage{enumerate}
\usepackage{enumitem}

\usepackage{dcolumn}
\usepackage{amsmath,amssymb}    
\usepackage{bm} 
\usepackage{hyperref}
\usepackage{latexsym}
\usepackage{color}
\usepackage{subfigure}
\usepackage{fancyhdr}
\def\beq{\begin{equation}}
\def\eeq{\end{equation}}

\def\beq{\begin{equation}}                           
\def\eeq{\end{equation}}                           
\def\bea{\begin{eqnarray}}                           
\def\eea{\end{eqnarray}}        

\begin{document}


\title{Ordering dynamics of self-propelled particles in an inhomogeneous medium}

\author{Rakesh Das}
\email[]{rakesh.das@bose.res.in}
\affiliation{S N Bose National Centre for Basic Sciences, Kolkata 700106}
\author{Shradha Mishra}
\affiliation{Department of Physics, Indian Institute of Technology (BHU), Varanasi 221005}
\author{Sanjay Puri}
\affiliation{School of Physical Sciences, Jawaharlal Nehru University, New Delhi 110067}


\begin{abstract}
Ordering dynamics of  self-propelled  particles in an inhomogeneous
medium in two-dimensions is studied. We write coarse-grained hydrodynamic equations of motion for 
density and  polarisation fields in the
presence of an external random disorder field, which is quenched in time. The strength of inhomogeneity is tuned 
from zero disorder (clean system) to large disorder.
In the clean system, the  polarisation 
field grows algebraically as  $L_{\rm  P} \sim t^{0.5}$. The density field 
does not show clean power-law growth; however, it  follows $L_{\rm \rho} \sim t^{0.8}$ approximately. 
In the {\it inhomogeneous} system, we find a disorder dependent growth.
For both the density and the  polarisation, growth slow down with increasing strength of disorder. 
The  polarisation shows a disorder dependent power-law growth 
$L_{\rm  P}(t,\Delta) \sim t^{1/\bar z_{\rm  P}(\Delta)}$  
for intermediate times.
At late times, there is a crossover to logarithmic growth $L_{\rm  P}(t,\Delta) \sim (\ln t)^{1/\varphi}$, where $\varphi$  
is a disorder independent exponent.  
Two-point correlation functions for the   polarisation shows dynamical scaling, but the density does not.\\ 
\begin{center}(Accepted in Europhysics Letters)\end{center}
\end{abstract}


\maketitle


Collective behaviour of self-propelled particles (SPPs) is observed in a wide variety of systems ranging from micron scales (as in a bacterial colony) to scales of the order of a few kilometers, {\it e.g.}, animal herds, bird flocks, etc.
\cite{animalgroup, helbing, feder, kuusela31, hubbard, rauch, benjacob,  harada, nedelec, schaller}.
Since the seminal work by Vicsek {\it et al.} \cite{vicsek}, collective behaviours of SPPs on homogeneous substrates are studied
extensively \cite{tonertu, tonertusr, srrmp, vicsekrev, chatepre, sppexp}. In these studies, the authors characterise different
varieties of orientationally ordered {\it steady states} in these systems. Recently, work has begun to study the effect of different kinds of inhomogeneity on the steady states of a collection of SPPs \cite{disorderst, bechinger}, as inhomogeneity is an inevitable fact of most natural systems.

The study of SPPs is complicated by the fact that the system settles into a non-equilibrium steady state (NESS). There have been very few studies \cite{chatepre} of the coarsening kinetics from a homogeneous initial state to this asymptotic NESS, though this is of great experimental interest. Previous studies of coarsening or domain growth have primarily focused upon systems approaching an equilibrium state \cite{ajbray1994,puribook}. The ordering dynamics of an assembly of SPPs, both in clean and inhomogeneous environments, is important to understand growth processes in many natural and granular systems. This is the problem we address in the present paper.
 
The SPPs are defined by their position and orientation (direction of velocity). Each particle moves along its orientation with a constant speed $v_0$ and tries to align with its neighbours. In addition to this, we introduce an inhomogeneous random field  
${\bf h}$ of fixed strength $\Delta$ and random orientation, but quenched in time. This random field locally aligns the orientation field along a preferred (but random) direction. Such a field may arise from physical inhomogeneities in the substrate, {\it e.g.}, pinning sites, impurities, obstacles, channels. The random field we introduce here is analogous to the random field in equilibrium spin systems \cite{im75,tn98}. We write the coarse-grained equations of motion for hydrodynamic variables: density and polarisation. We numerically solve these coupled nonlinear equations for different strengths of disorder. Starting from a random isotropic state, we observe coarsening of the density and the polarisation fields. Our primary focus in this study is the scaling behaviour and growth laws \cite{puribook} which characterise the emergence of the asymptotic NESS from the disordered state.

Before proceeding, we should stress that there does not as yet exist a clear understanding of the nature of the NESS in the case with substrate inhomogeneity. This problem definitely requires further study. Nevertheless, it is both useful and relevant to study the coarsening kinetics, even without a clear knowledge of the asymptotic state \cite{puribook}. As a matter of fact, a proper understanding of coarsening kinetics in the inhomogeneous system might also provide valuable information about the corresponding NESS.

In the absence of any inhomogeneity, {\it i.e.}, in a clean system, the  polarisation field grows algebraically with exponent $0.5$, while the density grows with an exponent close to $0.8$. However, the presence of inhomogeneities slows down the growth rate of the hydrodynamic fields in a complicated manner. For intermediate times, domains of the polarisation field follow a power-law growth with a disorder-dependent exponent. At late times, the polarisation field shows a crossover to logarithmic growth, and the logarithmic growth exponent does not depend on the disorder. For large disorder strength, the local polarisation remains pinned in the direction of the quenched random field. However, for the density field, we could not find corresponding unambiguous growth laws.


\begin{figure*}[t]
  \begin{center}
   \includegraphics[width=0.98\linewidth]{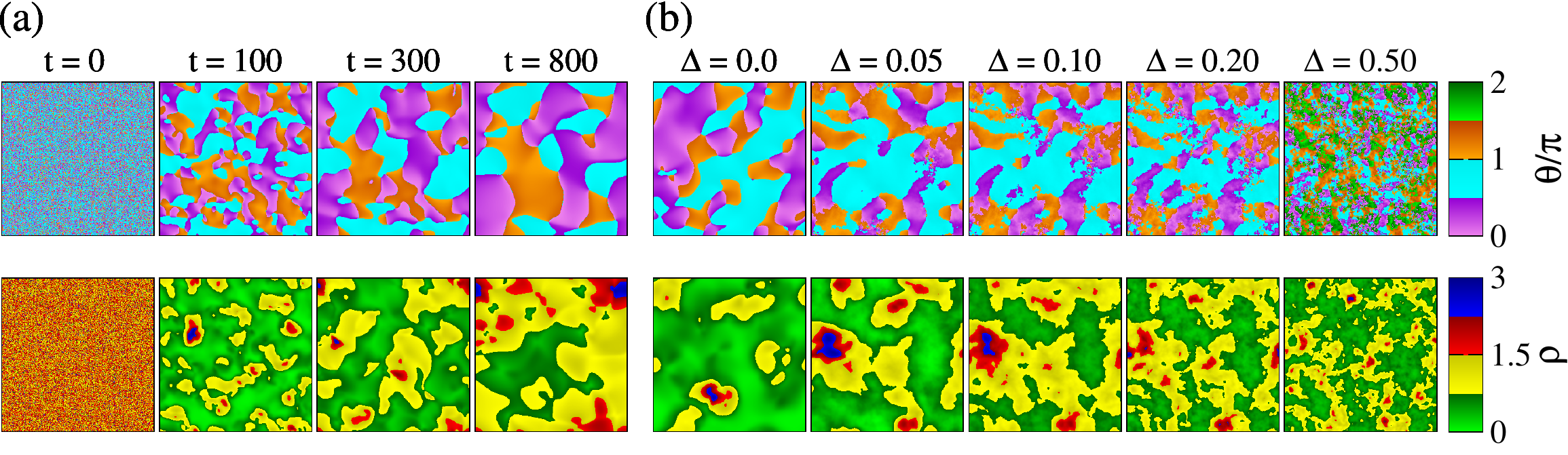}
\caption{(colour online)
Heat map of (upper panel) the orientation  $\theta({\bf r},t)=\tan^{-1}\left({\frac{{\rm P}_y({\bf r},t)}{{\rm P}_x({\bf r},t)}}\right)$, 
and (lower panel) the density. (a) shows the evolution of the respective fields with time in the clean 
system ($\Delta=0$). Starting with random orientation and uniform density at $t=0$, the system coarsens with time.
(b) is drawn for different disorder strengths at the same time ($t=1000$). Size of the ordered domains reduces with increasing 
strength of disorder.}
\label{fig1}
\end{center}
\end{figure*}


Let us first discuss our model. We consider a collection of SPPs of length $l$, moving on a two-dimensional substrate of friction coefficient $\chi$. Each particle is driven by an internal force $F$ acting along the long axis of the particle. The ratio of the force $F$ to the friction coefficient gives a constant  self-propulsion speed $v_0= F/\chi$ to each particle.
On time-scales large compared to the interaction time, and length scales 
much larger than the particle size, the dynamics of the system is governed by two 
hydrodynamic fields: density (which is conserved), 
and polarisation vector (which is a broken-symmetry variable in the ordered state). 
The ordered state is  also a moving state with mean velocity $v_0 {\bf P}$.
The dynamics of the system is characterised by the coupled equation of 
motion for the density and polarisation vector. The coarse-grained density equation is
\begin{equation}
\frac{\partial \rho}{\partial t} = -v_0 \nabla \cdot ({\bf P} \rho) + D_{\rho} \nabla^2 \rho .
\label{eq1}
\end{equation}
The corresponding polarisation equation is
\begin{align}
\frac{\partial {\bf  P}}{\partial t} & =[\alpha_1(\rho) - \alpha_2{\bf  P}\cdot {\bf  P}]{\bf  P} - \frac{v_0}{2 \rho}\nabla \rho  + \lambda_1 ({\bf  P} \cdot \nabla) {\bf  P}
                                    \notag \\
                                    & + \lambda_2 \nabla(|{\bf  P}|^2) + \lambda_3 {\bf  P}(\nabla \cdot {\bf  P}) + K \nabla^2 {\bf  P} + {\bf h}. 
\label{eq2}
\end{align}
The hydrodynamic eqs. (\ref{eq1}) and (\ref{eq2}) are of the same form as proposed on a phenomenological
basis  by Toner and Tu \cite{tonertu} to describe the physics of a collection of SPPs. Next we discuss the details of different terms in the above two equations.

In eq. (\ref{eq1}), $D_{\rho}$ represents diffusivity in the density field. Since the number of particles is conserved, we can express the R.H.S. of eq. (\ref{eq1}) as $-\nabla \cdot {\bf J}$, where the current ${\bf J}$  consists of terms ${\bf J}_D \propto \nabla \rho$ and an active current ${\bf J}_A \propto v_0 {\bf  P}\rho$. The active current arises because of the self-propelled nature of the particles.

The $\alpha$-terms on the R.H.S. of eq.~(\ref{eq2}) represent mean-field alignment in the system. For metric distance
interaction models, {\it e.g.}, the Vicsek model, these terms depend on the microscopic model parameters, 
{\it viz.}, mean density, noise strength etc. \cite{bertinnjop2009}. We choose $\alpha_1(\rho) = \frac{\rho}{\rho_c}-1$ and 
$\alpha_2=1$. Then the clean system (${\bf h}={\bf 0}$) shows a mean field transition from an isotropic disordered state 
with  ${\rm P}=0$ for mean density $\rho_0 < \rho_c$ to a homogeneous ordered state with ${\rm P}=\sqrt{\frac{\alpha_1(\rho_0)}{\alpha_2}}$ for $\rho_0 > \rho_c$. The $\nabla \rho$ term in eq. (\ref{eq2}) represents pressure in the system appearing because of density fluctuations. Here, ${\bf P}$ plays a dual role in the SPP system. First, it acts like a polarisation vector order parameter of same symmetry as a two-dimensional $XY$ model. Second, $v_0 {\bf P}$ is the flock velocity with which the density field is convected. Therefore, we choose same $v_0$ for the active current term in the density equation and the pressure term in the polarisation equation, because origin of both is the presence of non-zero self-propelled speed. As soon as we turn off $v_0$, the active current turns zero, and the density shows usual diffusive behaviour. Then we can ignore density fluctuations as well as the pressure term. However, in general they can be treated as two independent parameters. $\lambda$ terms  are the convective nonlinearities, present because of the absence of the Galilean invariance in the system. $K$ represents diffusivity in the  polarisation equation. 

To introduce inhomogeneity, {\it i.e.}, disorder in the system, a random-field term $\mathcal{F}_h=-{\bf h} \cdot {\bf P}$ is added 
in the `free energy'. This contributes the term  $-\frac{\delta \mathcal{F}_h}{\delta \bf  P} = {\bf h}$  in the  polarisation 
equation.  We should stress that such a term coupling to the polarisation field would not arise in the free 
energy of an equilibrium fluid, but may be realised in the context of the XY model where the polarisation vector is a spin variable. 
The random field is modeled as ${\bf h}({\bf r})=\Delta\left(\cos\psi({\bf r}),\sin\psi({\bf r})\right)$ where $\Delta$ represents
the disorder strength, and $\psi({\bf r})$ is a uniform random angle $\in [0, 2 \pi]$. We call the model defined by the hydrodynamic 
eqs. (\ref{eq1}) and (\ref{eq2}) as a `random field active model' (RFAM). This terminology originates from the well-known random-field 
Ising model (RFIM), which has received great attention in the literature on disordered systems  \cite{im75,tn98}. 
We are presently studying the phase diagram of the RFAM. However, a clear determination of this is complicated by the presence 
of long-lived metastable states. Apart from the RFAM, it is also natural to consider a random-bond active model (RBAM), 
where the average orientation  in the microscopic Vicsek model is weighted with `random bonds' for different 
neighbours. In this letter, we will focus on the RFAM.

For zero self-propelled speed, {\it i.e.}, $v_0=0$, eq. (\ref{eq1}) decouples from the  polarisation field and 
contains only the diffusion current. Hereafter, we refer to this as a  `zero-SPP model' (zero-SPPM). In the zero-SPPM, 
although it contains convective non-linearities, but coupling to density is only diffusive type. 
For $\Delta=0$, eqs. (\ref{eq1}) and (\ref{eq2}) reduce to the continuum equations introduced by Toner and Tu \cite{tonertu}, 
which represent the clean system.  While writing eqs. (\ref{eq1})-(\ref{eq2}), all lengths are rescaled by the interaction 
radius in the underlying microscopic model, and time by the microscopic interaction time.  In doing that all the coefficients 
(speed $v_0$, diffusivities $D_{\rho}$, $K$, non-linear coupling $\lambda$'s 
and field ${\bf h}$) are in  dimensionless units. 
Thus, eqs. (\ref{eq1})-(\ref{eq2}) are in
dimensionless units.

We should stress that the most general forms of eqs. (\ref{eq1})-(\ref{eq2}) also contain noise or ``thermal fluctuations''. 
For domain growth in non-active systems \cite{ajbray1994,puribook}, coarsening kinetics is dominated by a zero-noise (or zero-temperature) fixed point. This is because noise only affects the interfaces between domains, which become irrelevant compared to the divergent domain size \cite{po88}. In the present problem, we again have divergent (though different) domain scales for the density and polarisation fields, as we will see shortly. Therefore, it is reasonable to first study the zero-noise versions in eqs. (\ref{eq1})-(\ref{eq2}), as we do in the present paper. However, it is also important to undertake a study of the noisy model and confirm the irrelevance of noise.


We numerically solve eqs.~(\ref{eq1}) and (\ref{eq2}) for the hydrodynamic variables. The substrate size is $L \times L$ ($L=256, 512, 1024, 2048$) with periodic boundary conditions in both directions. An isotropic version of Euler's discretization scheme is used to approximate the partial derivatives appearing in the hydrodynamic equations of motion. In our numerical implementation, the first and second order derivatives for an arbitrary function $f({\bf r},t)$ are discretized as
\begin{eqnarray}
\frac{\partial f}{\partial t} &=& \frac{f(t+\Delta t)-f(t)}{\Delta t} ,\nonumber \\
\frac{\partial f}{\partial x} &=& \frac{f(x+\Delta x)-f(x-\Delta x)}{ 2 \Delta x}, \nonumber \\
\frac{\partial^2 f}{\partial x^2} &=& \frac{f(x+\Delta x)-2 f(x) + f(x-\Delta x)}{(\Delta x)^2},
\end{eqnarray}
where $\Delta t$ and $\Delta x$ are mesh sizes. While solving the equations, the field is specified on each grid point. Thus, we have a field of strength $\Delta$ and random orientation (which is quenched in time) at each grid point. The random angle is chosen from a uniform distribution in the range $[0, 2\pi]$. Our numerical scheme is convergent and stable for the chosen grid sizes $\Delta x =1.0$ and $\Delta t =0.1$.

We treat the parameters as phenomenological, and choose $-\lambda_1 = \lambda_2 = \lambda_3=0.5$, $D_{\rho} = 1$, $K = 1$ and $v_0 = 0.5$.  The above values of $\lambda$'s 
are chosen for simplicity. We checked that the homogeneous ordered steady state in the clean system is stable \cite{shradhapre} 
for the above choice of the parameters,  and that can become unstable for large $\lambda$'s. We start with a homogeneous 
isotropic disordered state with mean density $\rho_0 = 0.75$ and random  polarisation, and observe ordering dynamics 
for different strengths of the random field $\Delta \in [0,1]$. We assume the mean field critical density $\rho_c = 0.5$ for the 
clean system.


\begin{figure}[b]
  \begin{center}
  \includegraphics[width=0.98\linewidth]{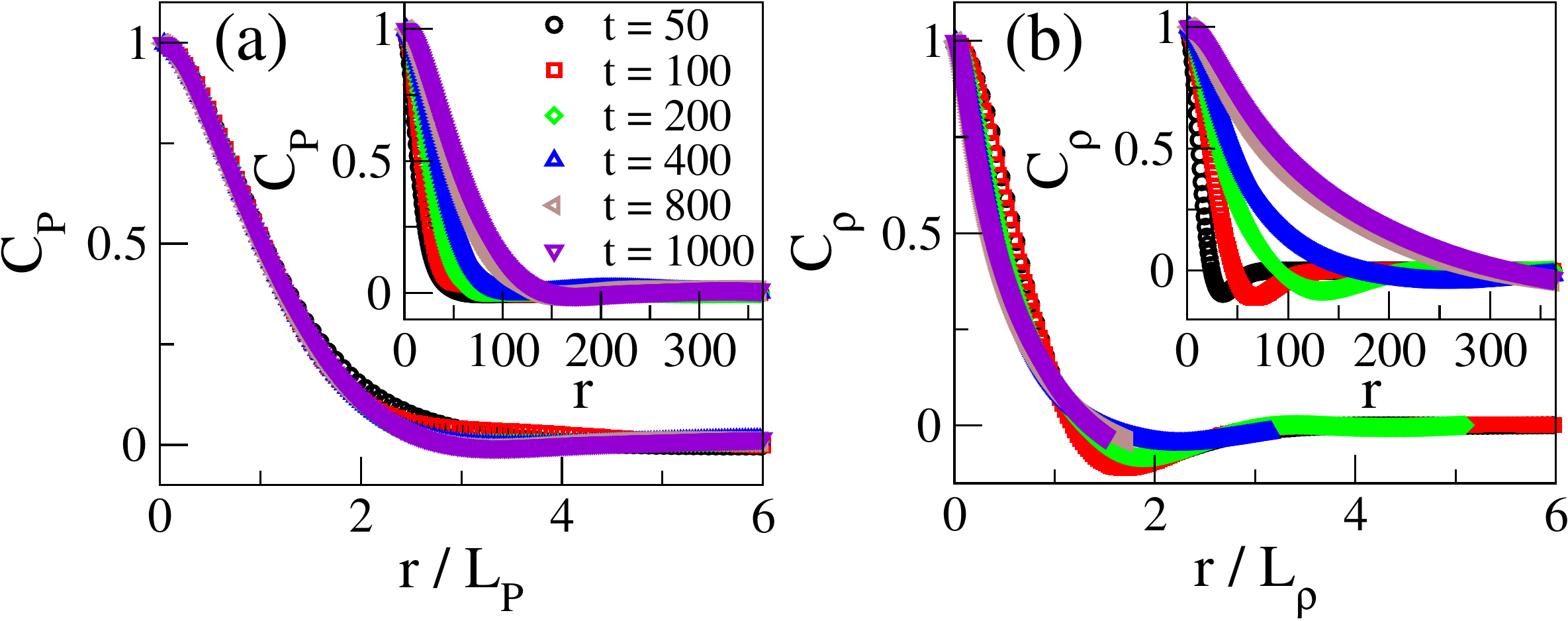}
\caption{(colour online)
The main figures show the two-point correlation functions for (a) the  polarisation and 
(b) the density in the clean system
($\Delta=0$), plotted with scaled distances. $C_{\rm  P}$ shows good collapse. The insets show  plots of correlation function 
versus distance for the respective fields at different times.}
\label{fig2}
\end{center}
\end{figure}


We first study the ordering dynamics of  the clean system, {\it i.e.}, $\Delta=0$. 
In fig. \ref{fig1}(a), we show snapshots of the orientation (upper panel) 
$\theta({\bf r},t)=\tan^{-1}\left({\frac{{\rm P}_y({\bf r},t)}{{\rm P}_x({\bf r},t)}}\right)$, and the density (lower panel) fields 
at different times. Starting from an initial isotropic state, high density domains with ordered  orientation 
emerge in the system, and the  size of these domains  increases with time. In the studies of domain growth in far-from equilibrium 
 systems \cite{ajbray1994,puribook}, the standard tool to characterise the evolution of morphologies is the equal-time
correlation function $C(r,t)$ of the order-parameter field. We use the same tool for the two fields ${\bf  P}({\bf r}, t)$ 
and $\rho({\bf r},t)$, which are relevant in the present context. We introduce the two-point correlation functions:
\begin{equation}
C_{\rm  P}(r, t) = \langle   {\bf  P}({\bf r_0}, t) \cdot   {\bf  P}({\bf r_0} +{\bf r}, t) \rangle_{\bf r_0},
\label{opcor}
\end{equation}
and
\begin{equation}
C_{\rm \rho}(r, t) = \langle \delta \rho({\bf r_0}, t) \delta \rho({\bf r_0} +{\bf r}, t) \rangle_{\bf r_0}.
\label{dencor} 
\end{equation}
 Here $\delta \rho $ represents fluctuation in the density from its  instantaneous local  mean value. 
Angular brackets denote spherical averaging (assuming isotropy), plus an average over space (${\bf r_0}$) and over 10 independent runs.

In fig. \ref{fig2} (a,b) insets, we show the correlation functions $C_{\rm  P}$ and $C_{\rm \rho}$ at different times for $\Delta=0$.
The data shows coarsening for both the fields, since the correlations increase with time. Characteristic lengths $L_{\rm  P}(t,\Delta)$ 
and $L_{\rm \rho}(t,\Delta)$ are defined as the distance over which the corresponding correlation functions fall to 0.5. 
In fig. \ref{fig2}(a,b) (main), we plot the correlation functions $C_{\rm  P}$ and $C_{\rm \rho}$, respectively, as a function of 
scaled distance $r/L_{\rm  P}$ and $r/L_{\rm \rho}$. We find nice scaling collapse for the polarisation, 
however, not for the density. Similar results are found for other disorder strengths (data not shown). 
The absence of dynamical scaling for the density correlation is consistent with the absence
of the single energy scale associated with the density growth dynamics \cite{ajbray1994}.

\begin{figure}[b]
  \begin{center}
  \includegraphics[width=0.98\linewidth]{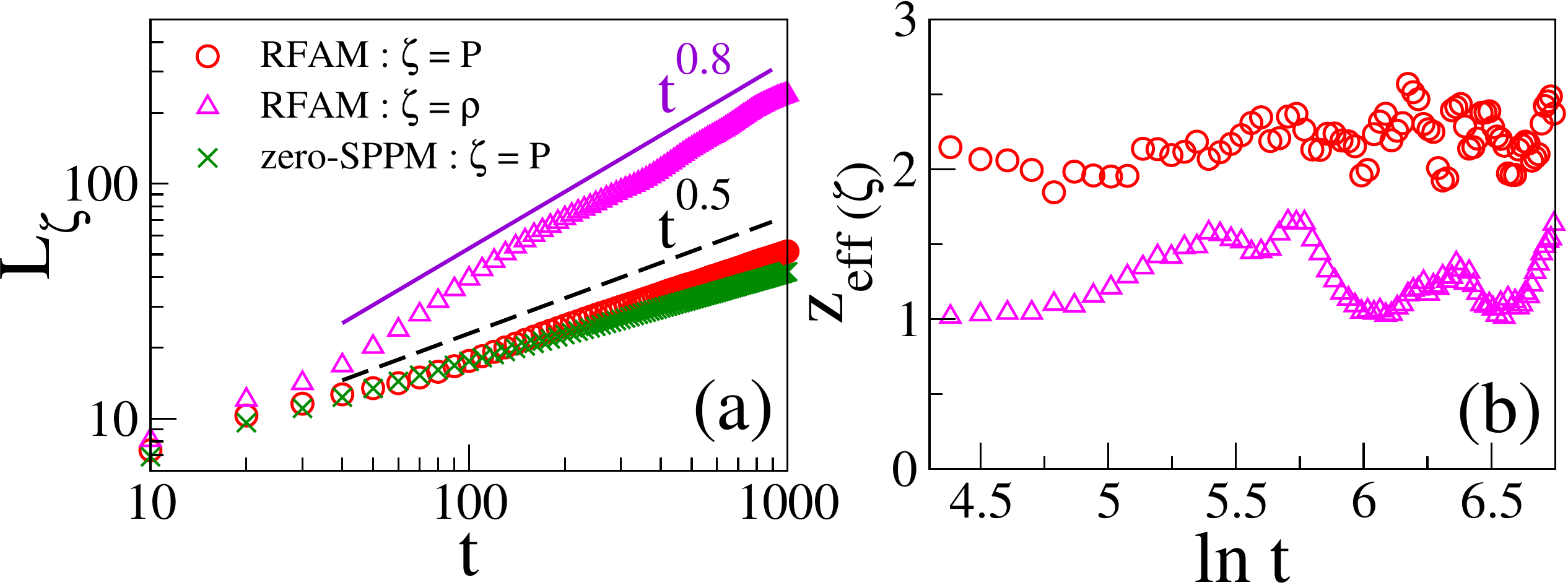}
\caption{(colour online) 
(a) Growth law of the hydrodynamic variables in the clean ($\Delta=0$) system. The self-propelled speed $v_0=0.5$ for the random field
active model (RFAM), whereas $v_0=0$ for the  zero-SPP model (zero-SPPM). The straight lines are drawn for the respectively 
indicated power-laws.
(b) Plot of effective growth exponent of the hydrodynamic fields versus time in the clean system for the RFAM.}
\label{fig3}
\end{center}
\end{figure}


In fig. \ref{fig3}(a), we show the time dependence of these length scales $L_{\rm \zeta}(t,0)$ where $\zeta \equiv ({\bf  P}, \rho)$.
We calculate the growth of the  polarisation
 field for two cases: (i) RFAM  with self-propelled speed $v_0=0.5$ and (ii)  zero-SPPM with $v_0=0.0$. 
For the clean system, we find that the characteristic length follows the similar growth law $L_{\rm  P}(t,0)\sim t^{0.5}$ 
for both the RFAM and the zero-SPPM.
The density shows usual diffusive growth for the zero-SPPM (data not shown).  
Although the data does not show clean power-law for the density, fig. \ref{fig3}(a) shows the  growth of the
characteristic length as $L_{\rm \rho}(t,0) \sim t^{0.8}$ for the RFAM in the clean system. 
Faster  growth of the density field in our study  is consistent with the previous study of self-propelled 
particles \cite{chatepre}. 
We define the algebraic growth law of the hydrodynamic fields in the clean system as $L_{\rm \zeta}(t,0) \sim t^{1/z_{\rm eff (\rm \zeta)}}$, 
where $z_{\rm eff (\rm \zeta)}$ is the effective growth exponent. 
In fig. \ref{fig3}(b), we show the variation of the effective growth exponent $z_{\rm eff (\zeta)}$ with time on log-linear scale 
for the two fields in the RFAM.
We find $z_{\rm eff (P)} \sim 2$ for almost two-decades, and $z_{\rm eff (\rho)} \sim 1.2$, 
when averaged over intermediate and late times, although it shows large oscillations.  These oscillations are not due to poor averaging, but rather an intrinsic feature of the density growth in active systems. These may arise due to the absence of a single energy scale for the density growth. 


\begin{figure}[b]
  \begin{center}
  \includegraphics[width=0.55\linewidth]{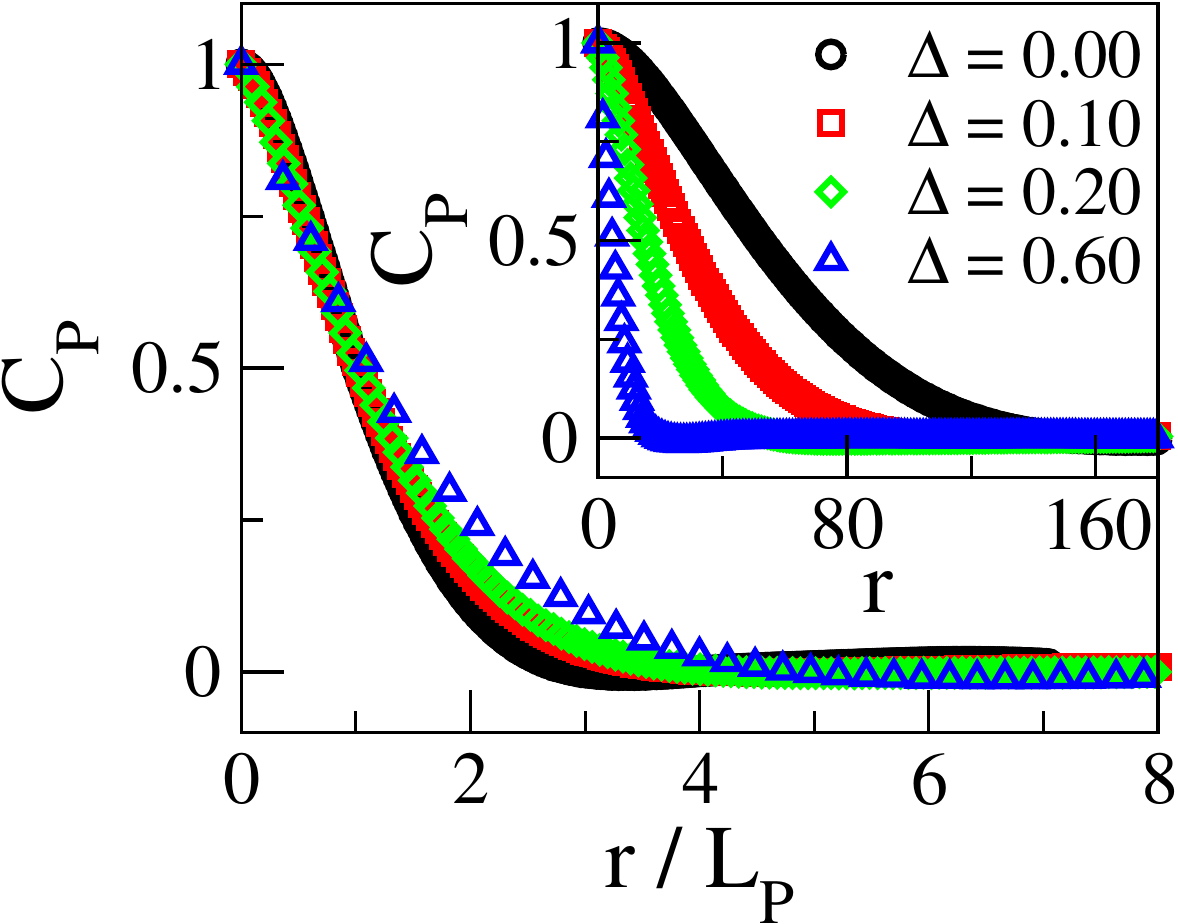}
\caption{(colour online)
Two-point correlation function for the  polarisation, drawn for different disorder strengths. 
The inset shows $C_{\rm  P}$ versus $r$ plot, and the main figure shows  scaling collapse of $C_{\rm  P}$ as a function 
of $r/L_{\rm  P}$.  Morphology of the polarisation field is approximately independent of disorder.} 
\label{fig4}
\end{center}
\end{figure}

\begin{figure}[b]
  \begin{center}
  \includegraphics[width=\linewidth]{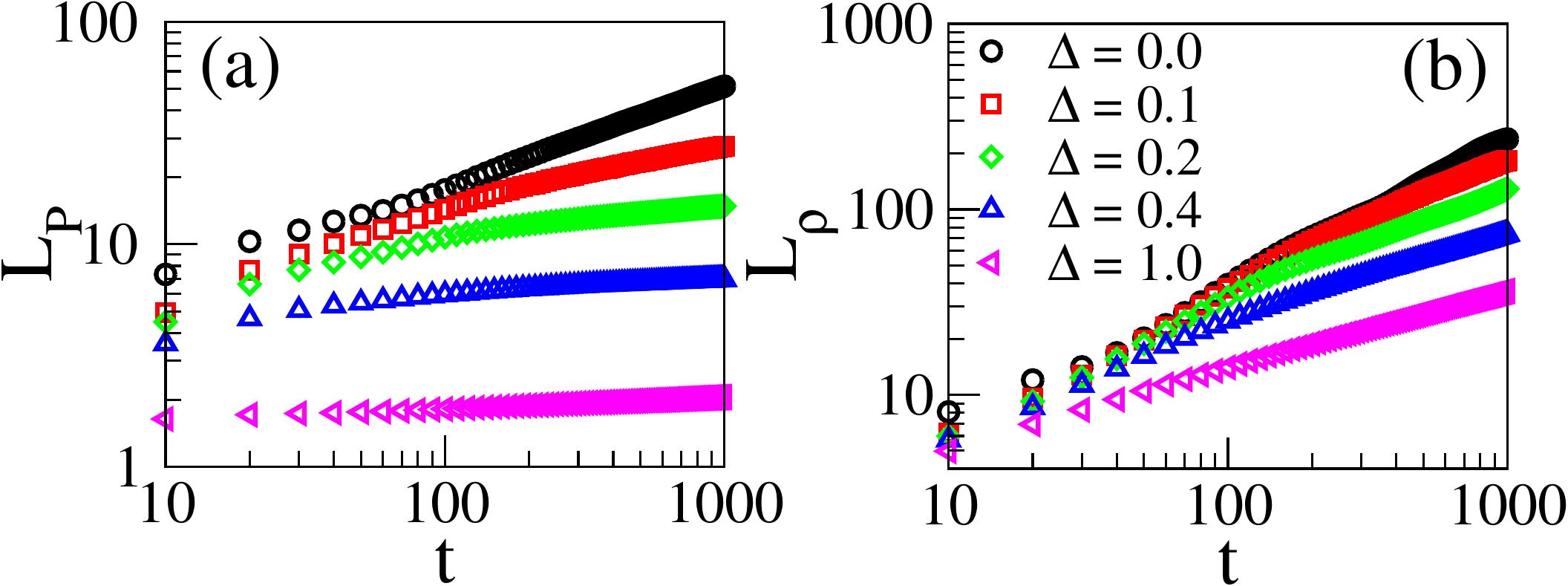}
\caption{(colour online)
Growth law of the field variables - (a) the  polarisation and 
(b) the density in the RFAM, drawn for different disorder
strengths. In the disordered environments, the growth deviates from the power-law at late times.} 
\label{fig5}
\end{center}
\end{figure}

Now we study the effect of disorder in the RFAM. 
In studies of domain growth, it has been found that random-field 
and random-bond disorder slows down the coarsening 
\cite{puriepl2017, huse1985prl, lai1988prb, puriparekh1992, paulpurireiger2004, purizannetti2011}.
This is attributed to the trapping of domain boundaries by sites of quenched disorder \cite{puriepl2017, huse1985prl, lai1988prb}. 
As most of the experimental 
systems contain disorder,  here we investigate the effect of random-field disorder on coarsening in the SPPs.
In fig. \ref{fig1}(b), we show snapshots of the orientation (upper panel) and the density (lower panel) 
at time $t=1000$ for different strength of disorder. We find that domain size decreases with increasing $\Delta$. The effect of 
inhomogeneity in the system is also inferred  from the  polarisation
 two-point correlation function shown in fig. \ref{fig4}(inset). 
Consequently, the characteristic lengths $L_{\rm  P,\rho}$ decrease with $\Delta$ as shown in figs. \ref{fig5}(a,b).
In fig. \ref{fig4}(main), we plot the two point correlation function $C_{\rm  P}$ vs. scaled distance $r/L_{\rm  P}$ 
for fixed time and different strengths of disorder $\Delta =0.0, 0.1, 0.2$ and $0.6$. We find good scaling collapse of the correlation 
functions. This suggests that the morphology of the polarisation field is approximately unaffected by disorder. However, this 
`super-universality' \cite{purichowparekh1991} does not extend to the density field which does not even show simple dynamical scaling.


\begin{figure}[t]
  \begin{center}
   \includegraphics[width=0.98\linewidth]{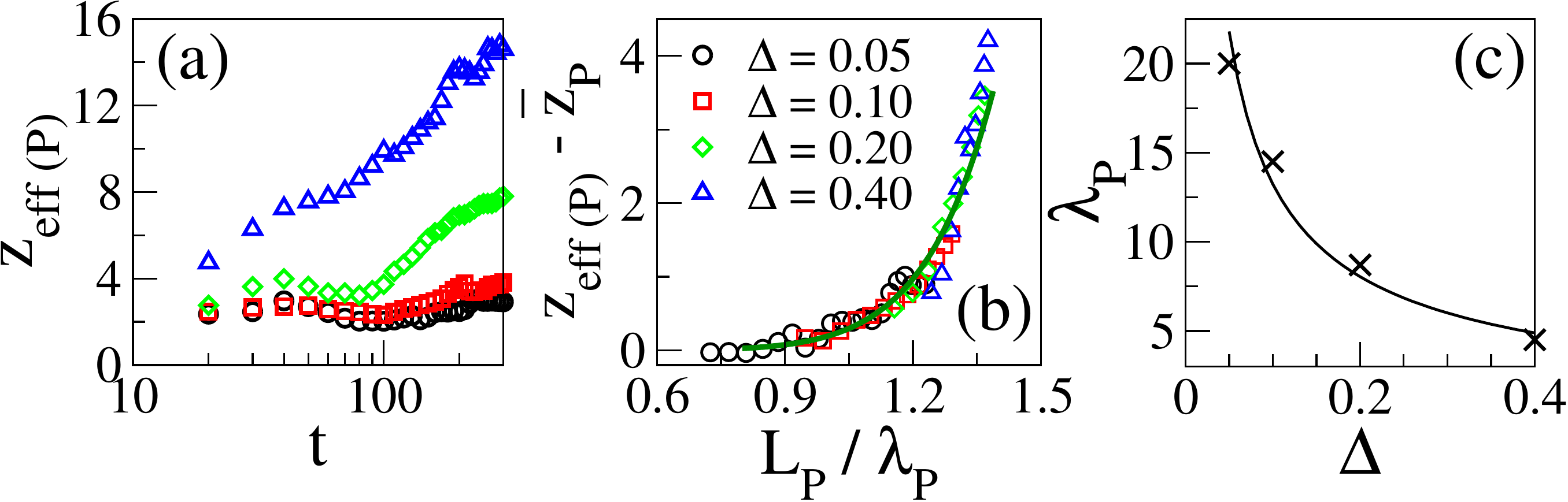}
\caption{(colour online)
(a) Time variation of the effective growth exponent of the  polarisation field in 
the RFAM, shown for different disorder strengths.
(b) The scaling collapse of $z_{\rm eff (P)} - \bar z_{\rm  P}$ versus $L_{\rm  P}/\lambda_{\rm  P}$. The best fit 
$z_{\rm eff (P)} - \bar z_{\rm  P} \simeq 
0.193(L_{\rm  P}/\lambda_{\rm  P})^{8.86}$ is shown by the solid line. (c) Disorder dependence of $\lambda_{\rm  P}$. The solid line shows a
power-law fit $\lambda_{\rm  P} \sim \Delta^{-0.72}$.}
\label{fig6}
\end{center}
\end{figure}

As stated before for the clean system,  $z_{\rm eff(P)}$ shows a mean value $\bar z_{\rm  P}(\Delta=0) \sim 2$ for 
 an extended range of time.  
In the RFAM, there is a preasymptotic regime with an effective exponent 
$\bar z_{\rm  P}(\Delta)$. As shown in fig. \ref{fig6}(a), $\bar z_{\rm  P}$ increases with $\Delta$.  
Also the preasymptotic regime decreases with increasing $\Delta$, and disappears for $\Delta>0.4$. 
Beyond the mean growth exponent regime, $z_{\rm eff(P)}(t,\Delta)$ increases sharply with time, 
that signifies pinning of the interfaces because of large disorder strength  \cite{purizannetti2011, equidis}. 

For the density field, we find $z_{\rm eff(\rho)}(t, 0) \sim 1.2$ (not clean power-law). 
 As we increase disorder strength, the effective growth exponent increases, but it does not show a clean power-law 
and fluctuates very much (data not shown).
Hereafter, we characterise the growth law in the presence of disorder for the {\it polarisation} field only. 

In the presence of disorder, we find a deviation from the power-law growth of the  polarisation field.
To analyse the effect of disorder, we use the method introduced by Corberi {\it et al.} \cite{equidis,purizannetti2011}. 
They propose the following scaling form for the growth law:
\begin{equation}
L(t, \Delta) \sim t^{1/z_{\rm eff}} = t^{1/{z}}F(\Delta/t^{\phi}).
\label{eqphi}
\end{equation}
Here $z_{\rm eff}(t,\Delta)$ represents the effective growth exponent, and $\phi$ is the crossover exponent. 
The scaling function $F(x)$ behaves as
\begin{equation}
F(x) \sim 
\begin{cases} 
\textrm{const.}, & \textrm{for } x \rightarrow 0, \\
x^{1/(z \phi )} \textrm{ } \ell \left( x^{-1/\phi} \right), & \textrm{for } x \rightarrow \infty,
\end{cases} 
\label{eqf}
\end{equation}
where $x=\Delta/t^{\phi}$.
For $\phi <0$, scaling form in eq. (\ref{eqphi}) shows a crossover from the power-law $L \sim t^{1/z}$ to an
asymptotic behaviour $L \sim \ell(t\Delta^{1/\lvert\phi\rvert})$. We evaluate the effective growth exponent
for the polarisation field using the relation $t=L^{z}G(L/\lambda)$ where the crossover length scale $\lambda = \Delta^{1/\phi z}$, 
and $G(y)=[F(x)]^{-z}$ with $y=L/\lambda$. Then the effective growth exponent is represented as a function of $y$ as
\begin{equation}
z_{\rm eff}(y) = \frac{\partial \ln t}{\partial \ln L} = z + \frac{\partial \ln G(y)}{\partial \ln y}.
\label{eqz}
\end{equation}
In fig. \ref{fig6}(a), we show the time dependence of $z_{\rm eff (P)}(t, \Delta)$ for $\Delta = 0.05, 0.1, 0.2$ and 
$0.4$. For the clean system, we find that $z_{\rm eff (P)}$ is close to $2$,  as shown in fig. \ref{fig3}(b). 
For non-zero $\Delta$, the plots show $z_{\rm eff (P)} \simeq \bar{z}_{\rm  P}$ for sufficient range of 
time. $\bar{z}_{\rm  P}$ is a disorder-dependent constant. This is followed by late time regime, where 
$z_{\rm eff (P)}$ is time-dependent. This  scenario seems to be a common feature of domain growth in disordered systems 
as shown in ref. \cite{equidis}. Hence we can write eqs. (\ref{eqphi}), (\ref{eqf}) and (\ref{eqz}) by replacing $z \rightarrow \bar{z}$.


Now we study the dependence of $z_{\rm eff}$ on $L$. From eq. (\ref{eqz}), we can say that
$z_{\rm eff}-\bar{z}$ only depends on $y=L/\lambda$. In fig. \ref{fig6}(b), we plot
$z_{\rm eff (P)}-\bar{z}_{\rm  P}$ vs. $L_{\rm  P}/\lambda_{\rm  P}$  for various disorder values. We choose 
different $\lambda_{\rm  P}$-values for different $\Delta$ to ensure the data collapse. 
The corresponding values of $\lambda_{\rm  P}$ and $\bar{z}_{\rm  P}$ for different $\Delta$ are listed in 
table \ref{table1}. The solid curve in  fig. \ref{fig6}(b) is the best-fit to the power-law form
\begin{equation}
z_{\rm eff}-\bar{z} = b y^{\varphi}
\label{eqfit}
\end{equation}
with $b=0.193$ and $\varphi = 8.86$. In fig. \ref{fig6}(c), we show the $\Delta$ dependence of 
 $\lambda_{\rm  P}$, which is fitted by $\lambda_{\rm  P} \sim \Delta^{-0.72}$.
The negative exponent implies that the disorder is indeed  a relevant scaling field.
From eq. (\ref{eqfit}) it is easy to confirm the logarithmic domain growth. The scaling
function $G(y)$ can be evaluated by
\begin{equation}
 \frac{\partial \ln G(y)}{\partial \ln y} = b y^{\varphi} \Rightarrow G(y) \sim \exp\left(\frac{b}{\varphi} y^{\varphi}\right). 
\label{eqgy}
\end{equation}
Substituting for $G(y)$ in eq. (\ref{eqz}) gives the asymptotic logarithmic growth form:
\begin{equation}
\frac{L}{\lambda} \simeq \left[\frac{\varphi}{b}\ln(t/\lambda^{\bar{z}})\right]^{1/\varphi}. 
\label{eql}
\end{equation}
 The exponent $\varphi$ has important physical significance in domain-growth studies as it measures how the 
trapping barriers scale with domain size. In our RFAM, we find $\varphi=8.86$. 



\begin{table}[b]
\caption{Parameters $\bar z_{\rm P}$ and $\lambda_{\rm P}$ in the RFAM with different $\Delta$ values.}
\centering
\begin{tabular}{|c|c|c|c|c|c|}
\hline
$\Delta$ & $0$ & $0.05$ & $0.10$ & $0.20$ & $0.40$ \\
\hline
$\bar z_{\rm  P}$ & $2.0$ & $2.06$ & $2.60$ & $3.40$ & $6.50$ \\
\hline
$\lambda_{\rm  P}$ & $\infty$ & $20.0$ & $14.50$ & $8.70$ & $4.50$ \\
\hline
\end{tabular}
\label{table1}
\end{table}


In summary, we  have studied ordering dynamics in a collection of
polar self-propelled particles in an inhomogeneous medium.
 We use a coarse-grained model, where  
inhomogeneity is introduced as an external disorder field, which is quenched in time and random in space.
The strength of disorder is tuned from $\Delta =0$ to $1.0$ and kept fixed during the evolution of the system.

When the system is quenched from a random isotropic state, both the density and the 
polarisation fields coarsen with time.
In the clean system, {\it i.e.}, $\Delta=0$, the polarisation field follows the power-law growth $L_{\rm  P}(t) \sim t^{0.5}$,
while the density field approximately grows as $L_{\rm \rho}(t) \sim t^{0.8}$. 
We find that the  polarisation shows dynamical scaling, whereas the density does not.
This indicates that the approach towards the  ordered state for the density field is no longer controlled
by a single energy scale associated with the cost of a domain wall.

The presence of disorder slows down the growth rate of the hydrodynamic 
fields. 
For intermediate time, domains of the  polarisation field follow a power-law growth 
$L_{\rm  P}(t, \Delta) \sim t^{1/\bar{z}_{\rm  P}(\Delta)}$ 
with a disorder-dependent exponent $\bar{z}_{\rm  P}(\Delta)$. 
At late times, the  polarisation field shows a crossover to logarithmic growth  $L_{\rm  P}(t, \Delta) \sim (\ln t)^{1/\varphi}$, 
where the exponent  $\varphi$  does not depend on disorder. We find the logarithmic exponent 
 is $\varphi=8.86$ 
for our two-dimensional RFAM. For large $\Delta$, the local  polarisation remains pinned in the direction of the quenched random field.
However, we could not find clean growth law for the density field.
The scaling function for $C_{\rm  P}(r,t)$ is approximately independent of disorder, showing that the morphology of the 
 polarisation field is relatively unaffected by disorder.  

In our present study, we find that the disorder plays an important role in the  phase ordering dynamics and scaling in a collection of SPPs. Our study provides novel insights on ordering dynamics in a collection of active polar particles in clean as well as disordered environments.  The disorder we introduce in our model is analogous to random fields introduced in usual spin systems. It would be interesting to study the effects of other kinds of disorder on ordering dynamics in active systems \cite{shradha2014phil,sdeyprl2012,amb2014natcomm}.

\begin{center}
{*****************}
\end{center}

R D thanks Manoranjan Kumar for useful discussion regarding the project. S M and S P would like to thank Sriram Ramaswamy for useful discussions at the beginning of the project. S M also thanks S N Bose National Centre for Basic Sciences, Kolkata for kind hospitality (where part of the work is done), and the DST-INDIA (INSPIRE) Research Award for partial financial support.


\end{document}